\documentclass[twocolumn,nofootinbib,
showpacs,pra,aps,floatfix]{revtex4}
\usepackage{graphicx}
\begin{document}
\def\la{\langle}
\def\ra{\rangle}
\def\ome{\omega}
\def\ome0{\omega_0}
\def\om0{\omega_0}
\def\Om{\Omega}
\def\vep{\varepsilon}
\def\wh{\widehat}
\def\P0{\wh{\cal P}_0}
\def\dt{\delta t}
\newcommand{\beq}{\begin{equation}}
\newcommand{\eeq}{\end{equation}}
\newcommand{\beqa}{\begin{eqnarray}}
\newcommand{\eeqa}{\end{eqnarray}}
\newcommand{\intf}{\int_{-\infty}^\infty}
\newcommand{\into}{\int_0^\infty}

\title{Ultra-fast propagation of Schr\"odinger waves in absorbing media}

\author {F. Delgado}

\author {J. G. Muga}

\author {A. Ruschhaupt}
\affiliation{Departamento de Qu\'{\i}mica-F\'{\i}sica,
UPV-EHU,\\
Apartado 644, 48080 Bilbao, Spain}

\begin{abstract}
We identify the characteristic times of the
evolution of a quantum wave generated by a point
source with a sharp onset in an absorbing medium. 
The ``traversal'' or ``B\"uttiker-Landauer'' time
(which grows linearly with the distance to the source)
for the Hermitian, non-absorbing case is substituted by  
three different characteristic quantities.  
One of them describes the arrival of a maximum of the density
calculated with respect to 
position, but the maximum with respect to time
for a given position 
becomes independent of the distance to the source and is given
by the particle's  ``survival time'' in the medium. 
This later effect, unlike the Hartman 
effect, occurs for injection frequencies under or above the 
cut-off, and for
arbitrarily large distances. A possible physical realization 
is proposed by illuminating a two-level atom with a detuned laser.  
\end{abstract}


\maketitle


As it was well understood by Brillouin and Sommerfeld long ago, 
certain wave features may travel at velocities exceeding 
$c$ in systems described by relativistic equations, without 
violating 
Einstein's causality \cite{SB}.
In non-relativistic equations one may 
similarly find 
``ultrafast'' phenomena which are 
subject to non-relativistic causality conditions \cite{MEDD02}.
In particular, 
the
Hartman effect \cite{Hartman62} 
has been  studied thoroughly
theoretically and experimentally
both for non-relativistic and relativistic equations 
\cite{HS89,LA90,Nimtz,LM94,BSM94,Chiao98,Ghose99,TQM,MEDD02}:
the 
peak of the transmitted packet in a collision of a
particle with a square
barrier emerges from the barrier edge at a time independent of the 
barrier width $d$ when the incident (average)
energy is below the barrier energy, 
and provided that the barrier is neither too thin nor too opaque.
This later large-$d$
limitation occurs because, for very opaque 
conditions, the wave packet components above the barrier dominate 
and lead to a linear-with-$d$ behaviour of the 
time of the peak \cite{Hartman62,BSM94}. 
For the barrier lengths where the Hartman effect holds  
one could in principle
transmit a peak
in a waveguide in such a way that it arrives 
simultaneously to receivers
located at different distances. We describe in this letter an even  
stronger effect for the Schr\"odinger equation: 
if the particles are sent by a point source with a sharp onset, 
or by a sudden opening of a shutter, to a region with an 
effective absorbing potential,   
the temporal maximum arrives at equal times at arbitrarily 
large positions, i.e., without the limitation of the Hartman 
effect to moderately opaque conditions.
Also at variance with the Hartman effect, this ultra-fast
propagation of a wave feature due to absorption 
also holds for injection energies
above the potential threshold.

Consider, for $x\ge 0$, the following 
Schr\"odinger equation 
with ``source'' boundary
conditions (all quantities are
dimensionless unless stated otherwise),
\beqa
&&i\frac{\partial \psi}{\partial t}=
-\frac{\partial^2 \psi}{\partial x^2}+(1-iV_1)\psi 
\label{wdim}
\\ 
&&\psi(x=0,t)=e^{-i\omega_0 t}\Theta(t)
\nonumber\\
&&
\psi(x,t)=0 \;\;\ x>0, t<0.
\nonumber
\eeqa
where $\omega_0$ is the injection frequency. This 
is a non-Hermitian generalization of the source with a sharp onset 
studied by several authors
before \cite{Stevens,Moretti,RMA91,BT,MB00,GVDM02}. 
The imaginary potential $-iV_1$ models the passage from the incident 
channel to other channels   
which are not
represented explicitly (a physical example is provided below). 
The dispersion relation corresponding to Eq. (\ref{wdim}) is
\beq
k=\sqrt{\omega-1+iV_1},\;\;\:{\rm Im}(k)\ge 0, 
\eeq
and the solution to Eq. (\ref{wdim}) is given by 
\beq
\psi(x,t)=\frac{i}{2\pi}\intf d\omega
\frac{e^{ikx-i\omega t}}
{\omega-\omega_0+i0},\label{solw}
\eeq
or, in the $k$-complex plane, 
as
\beqa
&&\psi(x,t)=\frac{ie^{-iVt}}{\pi}\int_{\Gamma_k}dk\, k\frac{e^{ikx-ik^2t}}
{k^2-k_0^2},
\label{solk}
\eeqa
where
\beqa
V&=&1-iV_1,
\\
k_0&=&k(\omega_0)=\sqrt{\omega_0-1+iV_1},\;\;\:{\rm Im}(k_0)\ge 0, 
\eeqa
and 
the countour $\Gamma_k$ goes from $-\infty$ to $\infty$ passing
above all singularities.  
Introducing a new integration variable
\beq
u=\frac{1+i}{\sqrt{2}}\sqrt{t}\left(k-\frac{x}{2t}\right),
\eeq
Eq. (\ref{solk}) takes the form
\beq
\!\psi(x,t)\!=\!\frac{ie^{-iVt+ix^2/4t}}{2\pi}\!\int_{\Gamma_u}\!\!du\,e^{-u^ 2}
\left(\frac{1}{u-u_0}+\frac{1}{u-u_0'}\right),
\label{solu}
\eeq
where
\beqa
u_0=\frac{1+i}{\sqrt{2}}\sqrt{t}\left(k_0-\frac{x}{2t}\right),
\cr
u_0'=\frac{1+i}{\sqrt{2}}\sqrt{t}\left(-k_0-\frac{x}{2t}\right),
\eeqa
and the contour $\Gamma_u$  goes from $-\infty$ to $\infty$
passing above the two simple poles.
Using the integral definition of the $w-$functions \cite{AS,MB00},
%
Eq. (\ref{solu}) is finally given by 
\beq
\psi(x,t)=\frac{e^{-iVt}e^{ix^2/4t}}{2}
\left[w(-u_0)+w(-u_0')\right].
\eeq
\begin{figure}
{\includegraphics[angle=-90,width=3.in]{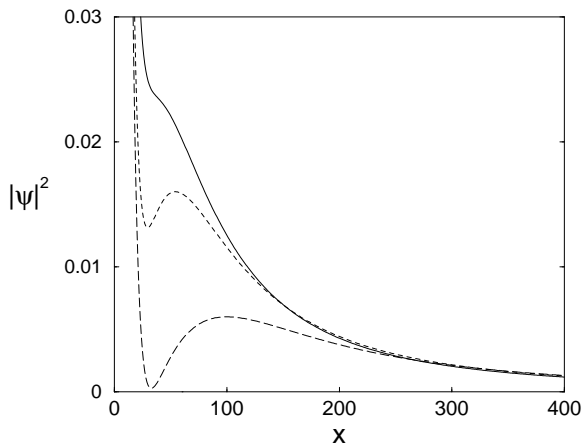}}
\caption[]{$|\psi|^2$ versus $x$ for three different values of $t$:
$t=250$ (solid line), $t=300$ (dashed line) and $t=500$ (long dashed 
line). We have chosen $\omega_0=0.99$ and $V_1=0.001$.} 
\label{f2}
\end{figure}
%
%

%
To interpret this exact solution in simple terms  
it is 
useful to find approximations. In particular,
the contour may be deformed along
the steepest descent path from the saddle 
at $x/2t$ in the $k$-plane (the origin in the $u$-plane).
The path crosses the pole at $k_0$ at a time
\beq
\tau_c\equiv x/2[{\rm Re}(k_{0})+{\rm Im}(k_{0})]. 
\eeq
%
Eq. (\ref{solu}) can be written as the sum of contributions
from the saddle point  
and the pole, 
\beqa
\psi(x,t)&=&\psi_p(x,t)+\psi_s(x,t)
\label{apro}\\
\psi_p(x,t)&=&e^{-i\omega_0 t+ik_0 x}
\Theta\left(t-\tau_c\right)
\\
\label{saddle}
\psi_s(x,t)&=&\frac{e^{-iVt}e^{i\frac{x^2}{4t}}}{2i\pi}
\left(\frac{1}{u_0}+\frac{1}{u_0'}\right)
\\
&=&\frac{e^{-iVt}e^{i\frac{x^2}{4t}}\tau(t/\pi)^{1/2}}
{(1+i)k_0(\tau^2+t^2)},
\eeqa
where
\beq
\tau\equiv \frac{x}{-2i k_0}.
\eeq
Alternatively, Eq. (\ref{apro}) may be obtained from the asymptotic
expansion of the 
$w$'s for large values of the modulus of their arguments $|u_0|$ and 
$|u_0'|$. They become large for
large $|k_0|,\, x$ or $t$, and also for very small $t$. 
Both have minima at $t=|\tau|$, with minimum values
$\{x[|k_0|\pm {\rm{Re}}(k_0)]\}^{1/2}$.
(A scale for the validity of the saddle 
plus pole approximation {\it for all} $t$ is
thus $x>[|k_0|-{\rm{Re}}(k_0)]^{-1}$).  
For injection frequencies below the cut-off  
the saddle contribution dominates up to exponentially large times as in 
the Hermitian case \cite{MB00}, so that $\tau_c$ 
is not of much significance whereas, above threshold, $\tau_c$ is a good 
scale for the arrival of 
the main signal. 
The discussion hereafter refers to the case $\omega_0<1$
(injection below cut-off)
unless stated otherwise.



In Figure \ref{f2} we can see the formation of the forerunner 
in a sequence of three snapshots of the density at three 
different instants and for a fixed value of $V_1$.  
(For increasing values of $V_1$ the spatial peak appears at larger $x$
and  it 
also 
takes a longer time to be formed.)  
From $d|\psi(t,x)|^2/dx=0$  
we may obtain the position $x(t)$ of the ``spatial'' maximum at time $t$.
This also 
defines (by inverting $x(t)$)
a function $\tau_S(x)$, namely, the time when this spatial maximum 
arrives at $x$.
One finds from Eq. (\ref{saddle})
that the spatial maximum in the large-$x$ region
is given by 
\beq
\tau_S=|\tau|,    
\eeq
a role played by the real $\tau$ in the absence of 
absorption \cite{VRS02}. 

A single real quantity $\tau(V_1=0)$ in the Hermitian case 
has been substituted, for a non-zero $V_1$, by  
three different quantities: the time for pole-cutting $\tau_c$, 
a complex $\tau$, and its modulus $|\tau|$, all of which tend to 
the ``B\"uttiker-Landauer'' traversal time $\tau(V_1=0)$ without
absorption \cite{BL82}.   

Similarly, we may fix $x$, calculate 
$d|\psi(x,t)|^2/dt=0$, and solve for $t$ to obtain a  
``temporal''
maximum, $\tau_T(x)$.

%
%
\begin{figure}
{\includegraphics[angle=-90,width=3.in]{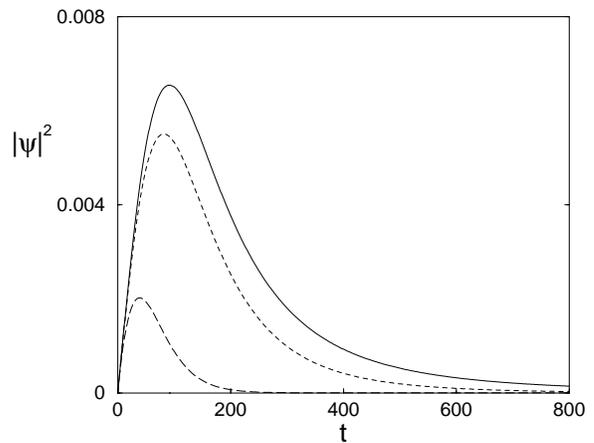}}
\caption[]{$|\psi|^2$ versus time for three different values of $V_1$:
$V_1=0$ (solid line), $V_1=0.001$ (dashed line) and $V_1=0.01$ (long dashed 
line). The position has been fixed to $x=100$ and $\omega_0=0.9$.} 
\label{f1}
\end{figure}
In Figure \ref{f1} we have plotted the density, $|\psi|^2$, versus time
for a 
large $x$, so that, in the scale used, 
the exact solution and the saddle aproximation $\psi_s$ are 
undistinguishable.
One clear effect 
of the complex potential is the decrease of the amplitude;
also, the peak arrives earlier when
$V_1$ increases. At variance with the Hermitian case, 
the time of arrival of the maximum is {\it not} proportional to 
$\tau$ ($\tau_T=\tau/\sqrt{3}$ when $V_1=0$).
The equation $d|\psi_s|^2/dt=0$ cannot be solved analytically 
for $V_1\ne 0$ in a generic case, 
so there is no explicit formula for $\tau_T$.  
Nevertheless, if $x\gg 2^{1/2}|k_0|/V_1$, one finds 
\beq
\tau_T\approx \frac{1}{2V_1},
\eeq
i.e., the temporal maximum coincides with the mean survival time 
of a particle immersed in  
the absorbing potential, and   
it is independent of $x$ and $\omega_0$, which is the 
most important result of this work. 
A time-frequency analysis \cite{Cohen95,MB00}
shows that the maximum 
is not tunnelling but it is dominated by frequencies above the cut-off, 
in particular by the frequency $1+(xV_1)^2$ 
corresponding to the ``classical'' 
velocity required to arrive at $x$ at time $\tau_T(x)$.  
In fact,
this effect is also present for injection frequencies above
the cut-off. 
It has, nevertheless, no purely classical explanation, in the
sense that any classical ensemble of particles injected 
at a constant rate into the absorbing 
medium with an arbitrary momentum distribution 
from $t=0$ on, would lead, for fixed $x$, to 
a monotonous increase
of the density up to the asymptotic, stationary value.

The important difference between temporal and spatial maxima 
may be understood from the  
scaling satisfied by the saddle term, 
%
%
\beq
|\psi_s(\eta x,\eta t)|^2=\frac{e^{-2tV_1(\eta-1)}}{\eta}|\psi_s(x,t)|^2.
\eeq
Since the exponential depends on $t$, but not on $x$,
the spatial maximum travels at constant velocity 
whereas, 
the temporal maximum does not, except for $V_1=0$. 

\begin{figure}
{\includegraphics[angle=-90,width=3.in]{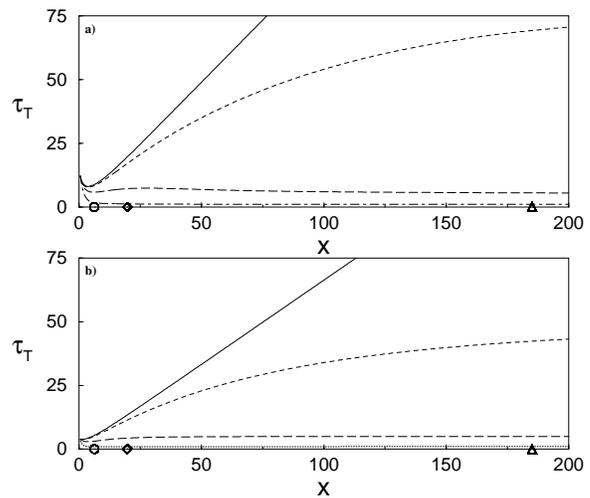}}
\caption[]{(a) Time of arrival of the temporal peak
versus $x$ for $V_1=0$ (solid
line), $V_1=0.001$ (dashed line), $V_1=0.01$ (long dashed line) and
$V_1=0.1$ (dotted-dashed line). The injection frequency is
$\omega_0=0.81$. (b)
Time of arrival of the peak versus $x$ for $V_1=0$ (solid
line), $V_1=0.01$ (dashed line), $V_1=0.1$ (long dashed line) and
$V_1=0.5$ (dots), and $k_0=0.9$.
The circle, rhomb, and triangle mark the value of $2^{1/2}|k_0|/V_1$
for $V_1=0.1, 0.01$, and $0.001$, respectively.} 
\label{f3}
\end{figure}

Some recent works have examined the time of arrival of the maximum at a
given position for the
source problem without absorption 
also for small values of $x$, where the pole-saddle 
approximation is not valid \cite{GVDM02,DMRGV}.  
For small $x$,
$\tau_T(x)$ presents in that case a basin with a minimum.  
In Fig. \ref{f3}a we have plotted this quantity 
versus $x$ for differents values
of the absorbing potential $V_1$.
Note that the basin 
dissappears by increasing the absorption. The most prominent 
feature of the curves though 
is their constant value $1/2V_1$ for large enough $x$,
instead of the linear dependence found without absorption.   
Unlike the Hartman effect, the arrival of the temporal peak
in an absorbing medium stays constant for arbitrarily large $x$.   
The effect is also present for other boundary conditions, in particular  
for the ``Moshinsky shutter'' boundary condition 
\cite{Moshinsky} corresponding to the 
initial, truncated-plane-wave state 
\beq
\label{initial}
\psi(x,t=0)=\frac{1}{\sqrt{2\pi}}e^{ik_0x}\Theta(-x),
\eeq
and the potential 
\beq
\label{poten}
V(x)=(1-iV_1)\Theta(x).
\eeq
Using the same techniques applied in \cite{DCM02}, we may calculate  
$\psi(x,t)$. 
The corresponding density is shown in Fig. \ref{f3}b, where 
the same characteristic time $1/2V_1$ 
which describes the arrival of the peak at large $x$
is found. 

%

A close physical realization of the Moshinki shutter problem with absorption 
may be based on a fluoresecence experiment where an atom 
is first prepared according to a truncated plane wave
in an internal state $|1\ra$,  
and then let evolve after a sudden shutter opening at $x=t=0$. 
The ``external'' region, $x>0$, is illuminated with a perpendicular laser. 
Let us assume a $\Lambda-$configuration for three relevant atomic
levels such
that the laser couples levels $|1\ra$ and $|2\ra$ whereas $|2\ra$ 
decays irreversibly by  
spontaneous photon emission to a ground state $|0\ra$. 
According to the quantum jump technique \cite{Hegerfeldt92} 
the amplitudes for 
levels $|1\ra$ and $|2\ra$ obey an effective Schr\"odinger 
equation with Hamiltonian (all quantities are now dimensional)
\cite{DEHM02,NEMH03,DEHM03}  
\begin{equation}
H= \hat{p}^2/2m + \frac{\hbar}{2} \left({0\atop \Omega\Theta(\hat{X})}
{\;\;\Omega\Theta(\hat{X})
    \atop \;\;-2\Delta-i\gamma} \right),
\label{hami2}
\end{equation}
where $\Omega$ is the Rabi frequency, $\Delta$ the detuning
between the laser frequency and the transition frequency $\omega_{12}$, 
$\gamma$ is the inverse life time of $|2\ra$,
and $\hat{p}$ the momentum operator 
for the initial atomic direction $X$.  For large detuning, 
$|2\Delta+i\gamma|>>\Omega$, a further reduction is possible to 
an even simpler effective Schr\"odinger equation
for the amplitude of level 1 with potential \cite{ChY91,OABSZ96,optim}
\beq
V(X)
=\frac{\hbar\Delta \Omega^2\Theta(X)}{4\Delta^2+\gamma^2}
-i\,\frac{\hbar\gamma\Omega^2\Theta(X)/2}{4\Delta^2+\gamma^2},  
\label{hami1}
\eeq 
which can be transformed by appropriate scaling 
to the form
of Eq. (\ref{poten}). 
We have solved numerically the full two-channel Schr\"odinger equation 
with the Hamiltonian of Eq. (\ref{hami2}) and the initial state 
of Eq. (\ref{initial}) 
for the $|1\ra-$component. We have found   
$\tau_T(x)=1/2V_1$ for arbitrarily large distances to the shutter,
in full agreement 
with the solution of the one-channel equation with absorbing 
potential, Eq. (\ref{wdim}).

\begin{acknowledgments}
We are grateful to I. L. Egusquiza, J. A. Damborenea, B. Navarro
and G. C. Hegerfeldt
for many discussions,  
and acknowledge support 
by Ministerio de Ciencia y Tecnolog\'\i a (BFM2000-0816-C03-03), 
and UPV-EHU (00039.310-13507/2001); 
AR also acknowledges a fellowship within the Postdoc-Programme of the 
German Academic Exchange
Service (DAAD). 
\end{acknowledgments}


\end{document}